\documentclass[aps,pra,twocolumn]{revtex4-2}

\usepackage{graphicx}
\usepackage{dcolumn}
\usepackage{bm}
\usepackage{physics}
\usepackage{float}
\usepackage{qcircuit}
\usepackage{amssymb}
\usepackage{hyperref}
\usepackage{subcaption}
\usepackage{multirow}
\usepackage{booktabs}

\hypersetup{
    colorlinks,
    citecolor=blue,
    linkcolor=blue,
    urlcolor=blue
}

\usepackage[whole]{bxcjkjatype}
\usepackage[normalem]{ulem}
\usepackage[dvipsnames]{xcolor}
\usepackage{soul}
\usepackage{pgfplots}
\pgfplotsset{compat=1.18}

\definecolor{electricindigo}{rgb}{0.44, 0.0, 1.0}
\definecolor{electricpurple}{rgb}{0.75, 0.0, 1.0}
\definecolor{electricultramarine}{rgb}{0.25, 0.0, 1.0}
\definecolor{americanrose}{rgb}{1.0, 0.01, 0.24}

\begin{document}

\title{Gradient Analysis of Barren Plateau in Parameterized Quantum Circuits with multi-qubit gates}

\author{Yuhan Yao}
\email{yao@biom.t.u-tokyo.ac.jp}
\author{Yoshihiko Hasegawa}%
\email{hasegawa@biom.t.u-tokyo.ac.jp}
\affiliation{%
 Department of Information and Communication Engineering,
Graduate School of Information Science and Technology,
The University of Tokyo, Tokyo 113-8656, Japan
}%
\date{\today}

\begin{abstract}
The emergence of the Barren Plateau phenomenon poses a significant challenge to quantum machine learning.
While most Barren Plateau analyses focus on single-qubit rotation gates, the gradient behavior of Parameterized Quantum Circuits built from multi-qubit gates remains largely unexplored.
In this work, we present a general theoretical framework for analyzing the gradient properties of Parameterized Quantum Circuits with multi-qubit gates.
Our method generalizes the direct computation framework, bypassing the Haar random assumption on parameters and enabling the calculation of the gradient expectation and variance.
We apply this framework to single-layer and deep-layer circuits, deriving analytical results that quantify how gradient variance is co-determined by the size of the multi-qubit gate and the number of qubits, layers, and effective parameters.
Numerical simulations validate our findings.
Our study provides a refined framework for analyzing and optimizing Parameterized Quantum Circuits with complex multi-qubit gates.

\end{abstract}

\maketitle

\textit{Introduction.---}
As quantum machine learning and variational quantum algorithms continue to advance, constructing and training larger-scale Parameterized Quantum Circuits (PQCs) has become a central focus of the field.
However, a significant obstacle arises as these circuits scale in the number of layers and qubits: the gradients required for optimization tend to vanish.
This phenomenon, known as the `Barren Plateau', severely impedes gradient-based optimizers, preventing convergence to meaningful solutions and posing a critical obstacle to achieving quantum advantage on NISQ devices.

Our analysis of the barren plateau phenomenon is situated within the theoretical study of PQCs, several studies \cite{yao2025directgradientcomputationbarren, McClean2018, Cerezo2021, Anshu2023intro, Leone2024, Uvarov2021gradient, Letcher2024gradient, napp2022gradient, Ragone2024lie, holmes_bp, Larocca_bp, sack_intro, du2022intro, shen2020intro, patti2021intro, wurtz2021intro, larocca2022intro, Russell2017intro, wiersema2020intro, Sciorilli2025intro, Wild2023intro, Palma2023intro, Caro2023intro, Jerbi2023intro, Wiersema2024intro} have employed theoretical analyses based on the principles of gradient expectation and variance.
To theoretically understand the origin of vanishing gradients, many seminal studies \cite{Grant_bp, yao2025, Ragone2024lie, McClean2018} have leveraged assumptions such as Haar-random integration or the formation of approximate t-designs.
These works established that, for deep, random PQCs measured with global observables, the gradient variance vanishes exponentially with the number of qubits.
In contrast, for shallow PQCs, studies \cite{Cerezo2021, Anshu2023intro} have demonstrated that using local observables can prevent this exponential decay, leading to lower variance and more stable convergence.
Other research lines have focused on estimating gradient variance by exploring its upper and lower bounds \cite{napp2022gradient, Uvarov2021gradient, Letcher2024gradient}, or have further elaborated on how circuit structure and the choice of observable jointly affect parameter contributions and the overall variance magnitude \cite{cerezo2024structure, Leone2024structure}.

Despite significant progress in analyzing barren plateaus, a substantial portion of existing theoretical works relies on the assumption that the circuits form approximate unitary $t$-designs (or imply Haar randomness).
While this assumption facilitates analytical derivations via the Weingarten formula, it attributes vanishing gradients to generic statistical properties of the Hilbert space.
However, this perspective overlooks the specific structural features of practical implementations, such as the Hardware Efficient Ansatz (HEA), which do not necessarily conform to such random ensembles.
While some analyses have moved towards these structured circuits, they have predominantly focused on PQCs constructed from single-qubit rotation gates. Consequently, the gradient behavior of more general PQCs built from multi-qubit rotation gates—whose rotations are generated by Pauli strings of local generator $s > 1$, such as $RXX(\theta)$ or $RXYZ(\theta)$—remains largely under-explored.
In this context, understanding how the interplay between specific circuit structures and multi-qubit generators impacts the gradient landscape becomes a pressing theoretical question.

In this Letter, we present a theoretical framework for analyzing the gradient properties of PQCs composed of arbitrary $s$-qubit gates ($s \ge 1$).
While the barren plateau phenomenon has been extensively studied, most existing analytical works restrict their attention to the special case of single-qubit unitaries ($s=1$).
This limitation is significant because it fails to capture the effects of multi-qubit entanglement blocks and overlooks how the gate size itself impacts gradient scaling.
To address this, we generalize the direct gradient computation technique from \cite{yao2025directgradientcomputationbarren} to the general $s$-qubit setting, enabling an exact analysis without relying on the Haar random assumption.
Our central finding quantifies the gradient distribution in both single-layer and deep circuit regimes, demonstrating that the variance is governed precisely by the system size $n$, circuit layer $l$, the number of effective parameters $N_{\mathrm{eff}}$, and, crucially, the multi-qubit gate size $s$.
This analysis offers new theoretical insights into understanding and navigating the barren plateau phenomenon in this broader class of PQCs.

\textit{Preliminaries.---}
Parametrized Quantum Circuits (PQCs) are a central component of modern quantum machine learning and variational quantum algorithms.
These quantum circuits are designed with tunable parameters ($\boldsymbol{\theta}$) that can be optimized to solve a variety of computational problems.
Several key elements fundamentally characterize PQCs:

\begin{table}[h]
    \centering
    \label{tab: pqc_components}
    \begin{tabular}{c c l}
        \toprule
        \textbf{Category} & \textbf{Symbol} & \textbf{Description} \\
        \midrule
        \multirow{4}{*}{\shortstack{Circuit\\Structure}} 
          & $n$ & Number of Circuit Qubits \\
          & $l$ & Number of Circuit Layers \\
          & $N_\mathrm{eff}$ & Number of Effective parameters \\
          & $s$ & Size of Multi-qubit Gate \\
        \midrule
        \multirow{3}{*}{\shortstack{Quantum\\Elements}} 
          & $U(\boldsymbol{\theta})$ & Circuit Structure \\
          & $\ket{\mathbf{init}}$ & Initial Quantum State \\
          & $O$ & Observable \\
        \midrule
        \multirow{2}{*}{\shortstack{Optimization\\Elements}} 
          & $\mathcal{L}(\boldsymbol{\theta})$ & Loss Function \\
          & $\partial_k \mathcal{L}$ & Loss Function's Gradient \\
        \bottomrule
    \end{tabular}
    \caption{\raggedright Key Components in Parametrized Quantum Circuits (PQCs).}
\end{table}

The full circuit structure $U(\boldsymbol{\theta})$ is constructed from alternating layers of parameterized gates and non-parameterized fixed gates.
It can be formally expressed as a product of $l$ blocks:

\begin{align}
U(\boldsymbol{\theta})=\prod_{i=1}^{l}U_{i}(\theta_{i})W_{i},
\end{align}
where $U_{i}(\boldsymbol{\theta}_{i})$ represents the parameterized quantum gates, such as the rotation gate, which contain the set of parameters $\boldsymbol{\theta}_{i}$ that are updated during the optimization process.
$W_{i}$ represents the fixed gates, such as $CX$ or $CZ$, which are essential for introducing entanglement within the circuit layers.

The fundamental building block for a parameterized rotation is the gate $RP(\theta)$, which is defined as:
\begin{align}
\label{eq: rotation_gate_define}
RP(\theta)=e^{-i\frac{\theta}{2}P}=\cos\frac{\theta}{2}I+i\sin\frac{\theta}{2}P,
\end{align}
where $P$ is an $s$-qubit Pauli operator, $P\in \{I,X,Y,Z\}^{\otimes s}$.
The total number of independent parameters in the circuit $|\boldsymbol{\theta}|$ is determined by the hardware and ansatz design, typically scaling with the number of qubits, layers, and the size of the multi-qubit gate.
For the specific PQCs considered here, which contains $l$ layers, with each layer contributing $\frac{n}{s}$ parameters, the total parameter count is given by:
\begin{align}
\label{eq: parameter_number}
|\boldsymbol{\theta}|=\frac{nl}{s}\geq N_\mathrm{eff}.
\end{align}

The initial state defines a quantum state $\ket{\mathbf{init}}$, and the circuit ends with measurements to output the expectation values of observables, which are used for optimization.
The loss function, calculated from the measured outputs and serving as the primary metric for optimization, can be expressed as follows:
\begin{align}
\label{eq: loss_function}
\mathcal{L}(\boldsymbol{\theta}) &= \bra{\mathbf{init}}U(\boldsymbol{\theta})^\dagger H U(\boldsymbol{\theta})\ket{\mathbf{init}}\notag\\
&=\text{Tr}\left(\rho U(\boldsymbol{\theta})^\dagger O U(\boldsymbol{\theta}) \right),
\end{align}
where \(\rho\) denotes \(\ket{\mathbf{init}}\bra{\mathbf{init}}\).
Gradient descent is the core of optimization. 
It iteratively updates parameters based on gradient calculations to progressively approach the optimal solution for specific application goals.

We now proceed to analyze the gradient expectation.
Following the derivation in Ref.~\cite{McClean2018}, the expression can be given as:
\begin{align}
    \mathrm{E}\left[\partial_k \mathcal{L}\right]&=\frac{i}{2}\mathrm{E}\left[\text{Tr} \left(O_+ \left[\rho_-, P_k\right]\right)\right],\\
    \mathrm{E}\left[\left(\partial_{k}\mathcal{L}\right)^2\right]&=-\frac{1}{4}\mathrm{E}\left[\text{Tr} \left(O_+  \left[\rho_-, P_k\right] O_+  \left[\rho_-, P_k\right]\right)\right],
\end{align}
where the $O_+$ term is defined as $O_+ = U_+^\dagger O U_+$, the $\rho_-$ term represents $\rho_- = U_- \rho U_-^\dagger $.
So the gradient variance can be given by 
\begin{align}
\label{eq: gradient_variance}
\mathrm{Var}\left[\partial_k \mathcal{L}\right] =\mathrm{E}\left[\left(\partial_{k}\mathcal{L}\right)^2\right] - \mathrm{E}\left[\partial_k \mathcal{L}\right]^2.
\end{align}

To obtain concrete analytical results, we must derive the first and second moments of $U(\boldsymbol{\theta})$.
We explore a method \cite{yao2025directgradientcomputationbarren} to compute these moments, which is generalizable from $s=1$ to arbitrary $s$-qubit gates.
This approach successfully circumvents the Weingarten formula, traditionally necessary for the second-moment calculation.

\textit{Results.---}First, we consider a single arbitrary real parameter $\theta$.
When all unitary gates in the PQCs are rotation gates, which are defined in Eq.~\eqref{eq: rotation_gate_define}, we obtain the following analytical result for the first moment:
\begin{align}
    &\mathrm{E}[U(\theta)^\dagger AU(\theta)]\notag\\
=&\mathrm{E}_{\theta,P}\left[\left(\cos\frac{\theta}{2}I+i\sin\frac{\theta}{2}P\right) A\left(\cos\frac{\theta}{2}I-i\sin\frac{\theta}{2}P\right)\right]\notag\\
=&\mathrm{E}_{\theta,P}\left[\cos^2\frac{\theta}{2}A+i\sin\frac{\theta}{2}\cos\frac{\theta}{2}(PA-AP)+\sin^2\frac{\theta}{2}PAP\right]\notag\\
=&\frac{1}{2}\mathrm{E}_{P}\left[A+PAP\right]\notag\\
=&\frac{A}{2}+\frac{1}{2|P|}\sum_{P}PAP.
\end{align}

We use the result of $\cos^2\frac{\theta}{2}=\sin^2\frac{\theta}{2}=\frac{1}{2}$ and $\sin\frac{\theta}{2}\cos\frac{\theta}{2}=0$, which can be find in \cite{SM}.
Similarly, the second moment can be expressed in a similar form (see Appendix B for more details).
Therefore, the problem of calculating the first and second moments of $U(\theta)$ is reduced to evaluating the $\sum PAP$ term for different combinations of Pauli operators.

Further analysis indicates that for specific combinations of Pauli operators, such as $P \in I^{\otimes a}\otimes\{I, X, Y, Z\}^{\otimes s}\otimes I^{\otimes b}$, where $a+s+b=n$.
We can derive a general expression for $\sum_P PAP$ that is independent of the specific operator choice, namely:
\begin{align}
    \sum_{P}PAP&=2^s \text{Tr}_s(A)\otimes I_s^{\otimes s},\\
    \sum_PPAPBPCP&=2^s\text{Tr}_s(AC)\otimes I_s^{\otimes s} B+\varepsilon,
\end{align}
where $\text{Tr}\{\varepsilon\}=0$.
A similar result also exists for the second moment.
This expression depends only on the partial trace operation of the matrix and is independent of the specific form of the Pauli operators.

Based on the single-parameter analysis above, we extend the parameter range to $|\boldsymbol{\theta}|=\frac{nl}{s}$, as defined in Eq. \eqref{eq: parameter_number}.
The detailed derivation process is provided in the \cite{SM}.
From this, we derive the final analytical expressions for the first and second moments of $U(\boldsymbol{\theta})$.
The expectation value (first moment) of a single operator $A$ is denoted as $\mathrm{E}[U(\boldsymbol{\theta})^\dagger AU(\boldsymbol{\theta})]$.
Similarly, the expectation of the second moment $\mathrm{E}\left[U(\boldsymbol{\theta})^\dagger A U(\boldsymbol{\theta})BU(\boldsymbol{\theta})^\dagger CU(\boldsymbol{\theta}) \right]$ is necessary for calculating the variance.
These expectation values are given by:
\begin{widetext}
\begin{align}
\label{eq: first_moment}
    \mathrm{E}[U(\boldsymbol{\theta})^\dagger AU(\boldsymbol{\theta})]&=\left(\frac{1}{2}\right)^{n}\sum_{\sigma\in\mathcal{P}\left(\frac{n}{s}\right)}\left(2^{-\frac{n}{s}}3^{|\sigma|}\right)^{l-1} 2^{-s|\sigma|}\text{Tr}_\sigma (A)\otimes I_\sigma^{\otimes s|\sigma|},\\
    \label{eq: second_moment} \mathrm{E}\left[U(\boldsymbol{\theta})^\dagger A U(\boldsymbol{\theta})BU(\boldsymbol{\theta})^\dagger CU(\boldsymbol{\theta}) \right]&=\left(\frac{3}{8}\right)^{\frac{n}{s}}\sum_{\sigma\in\mathcal{P}\left(\frac{n}{s}\right)}\left(8^{-n}3^{|\sigma|+\frac{n}{s}}\right)^{l-1} 2^{-s|\sigma|}\text{Tr}_\sigma \left(AC\right)\otimes I_\sigma^{\otimes s|\sigma|} B+O(8^{-n}).
\end{align}
\end{widetext}
The result of Eq. \eqref{eq: first_moment} and Eq. \eqref{eq: second_moment} constitutes the core contribution of this paper.

\textit{Analysis of Single-Layer PQCs.---}
Our analysis explores the properties of gradients in single-layer PQCs.
The method involves averaging over the $s$-qubit Pauli group, extending from the single-qubit Pauli operators $\{I, X, Y, Z\}$ to the tensor-product basis $\{I, X, Y, Z\}^{\otimes s}$.

Single-layer PQCs represent a special case of Eq. \eqref{eq: first_moment} and Eq. \eqref{eq: second_moment} when $l=1$.
Using these identities, we can derive the expectation values for operators under this averaging, which models the behavior of a random circuit ensemble.
We now apply this framework to the cost function gradient, $\partial_k \mathcal{L}$.
In this framework, it is essential to distinguish between `effective' and `ineffective' parameters.
The gradient variance for ineffective parameters vanishes as $\mathrm{Var}[\partial_k \mathcal{L}] = 0$.
Conversely, for the effective parameters, the variance is found to be $\mathrm{Var}[\partial_k \mathcal{L}]=F(P) G(P)^{N_\mathrm{eff}-1}$, where $F(P)$ and $G(P)$ are functions specified in Appendix C and \cite{SM}.
By combining these findings, we obtain a concise result for the expectation and variance of the gradient with respect to a randomly chosen parameter:
\begin{align}
\mathrm{E}[\partial_k \mathcal{L}] &= 0, \\
\label{eq: single_layer_variance}
\mathrm{Var}[\partial_k \mathcal{L}] &= \frac{sN_\mathrm{eff}}{n} F(P) G(P)^{N_\mathrm{eff}-1}.
\end{align}
Equation (\ref{eq: single_layer_variance}) provides a unified framework to characterize the gradient scaling in single-layer PQCs.
Notably, our result suggests that the trainability of the circuit is primarily governed by the number of effective parameters $N_{\mathrm{eff}}$ and the gate size $s$, rather than the conventional distinction between local and global observables.
By focusing on $N_{\mathrm{eff}}$, this framework bypasses the need for complex geometric analysis of the observable's locality, offering a more direct and quantifiable metric to predict the onset of barren plateaus.

\begin{figure}[t]
    \includegraphics[width=\linewidth]{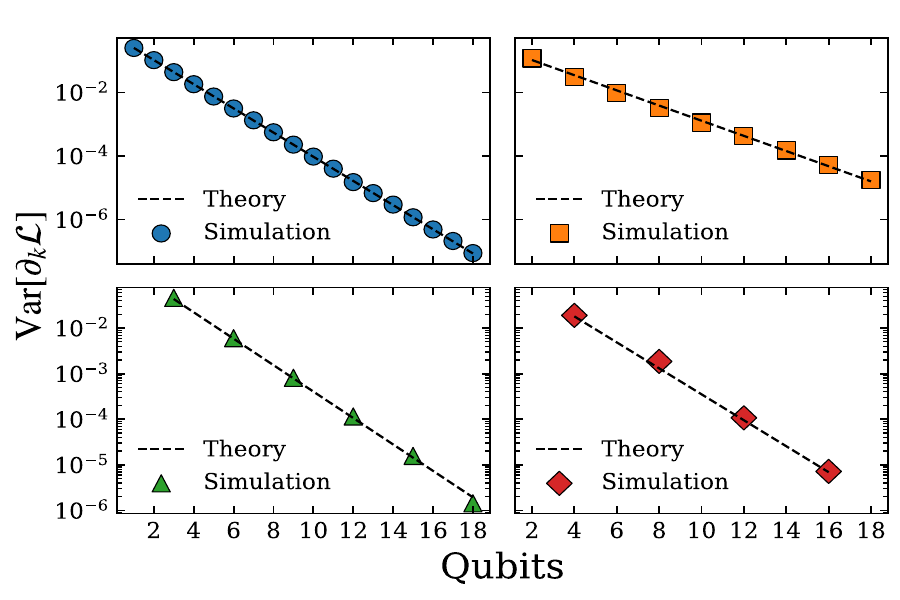}
    \caption{\raggedright Comparison of different settings of the theoretical analysis in single-layer PQCs from Eq. \eqref{eq: single_layer_variance} using numerical simulations. The plots compare the theoretical predictions (dashed lines) with numerical simulation results (scatter) under the condition $N_\mathrm{eff} = \frac{n}{s}$. The four subplots correspond to different values of the $s$-qubit gate: $s=1$ with $ \frac{1}{4} (\frac{5}{12})^{n-1}$ (top-left), $s=2$ with $\frac{5}{48} (\frac{1}{3})^{\frac{n}{2}-1}$ (top-right), $s=3$ with $\frac{25}{576} (\frac{17}{126})^{\frac{n}{3}-1}$ (bottom-left), and $s=4$ with $\frac{125}{6912} (\frac{37}{510})^{\frac{n}{4}-1}$ (bottom-right).}
    \label{fig: single_layer}
\end{figure}

Next, we perform numerical simulations to validate our preceding theoretical result.
Under the condition $N_\mathrm{eff}=\frac{n}{s}$, we examine cases where the $s$-qubit gate ranges from $1$ to $3$.
Figure~\ref{fig: single_layer} plots the numerical results from these simulations against the theoretical values predicted by our analysis of Eq. \eqref{eq: single_layer_variance}.

\begin{figure}[h]
    \centering
    \includegraphics[width=0.9\linewidth]{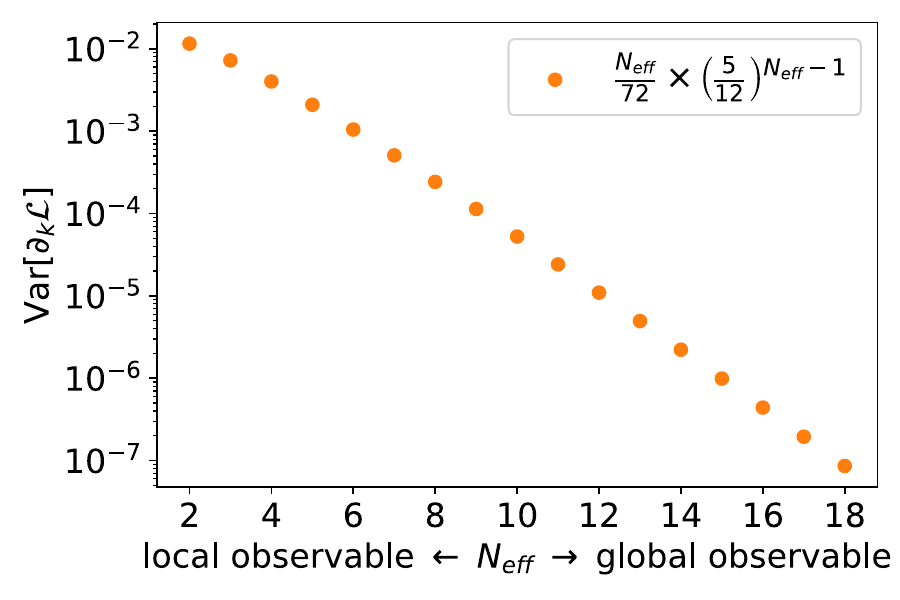}
    \caption{\raggedright Behavior of the gradient variance in single-layer PQCs as a function of $N_\mathrm{eff}$. The x-axis illustrates the transition from local observables (low $N_\mathrm{eff}$) to global observables (high $N_\mathrm{eff}$). The circuit qubit and generator are fixed at $n=18$ and $s=1$.}
    \label{fig: single_observable}
\end{figure}

Furthermore, Figure~\ref {fig: single_observable} illustrates the behavior of the gradient variance as a function of $N_\mathrm{eff}$, with $n=18$ and $s=1$.
We clearly observe that gradient variance exhibits an exponential decay as the number of effective parameters ($N_\mathrm{eff}$) increases.
This trend aligns with the findings in Ref.~\cite{Cerezo2021}, which noted that gradient variance performs better for local observables (corresponding to small $N_\mathrm{eff}$) and worse for global observables (corresponding to large $N_\mathrm{eff}$).
Building upon the outstanding work in Ref.~\cite{Cerezo2021}, which identified distinct behaviors for `local' and `global' observables, our analysis provides a quantitative perspective.
We show that these two categories correspond to the regimes of small and large $N_\mathrm{eff}$, respectively.
While the original distinction between `local' and `global' is insightful, its boundary can be imprecise.
We propose that the number of effective parameters serves as a more quantitative and unifying parameter, offering a clearer framework that extends this concept and unifies both regimes.

\textit{Analysis of Deep-Layer PQCs.---}
Following a similar approach, we now apply this analytical framework to the cost function gradient for deep-layer PQCs.
When the number of layers $l$ is sufficiently large, our analysis is likewise based on Eq. \eqref{eq: first_moment} and Eq. \eqref{eq: second_moment}.
Since the gradient expectation $\mathrm{E}[\partial_k \mathcal{L}]$ is approximately zero, our analysis concentrates on the variance $\mathrm{Var}[\partial_k \mathcal{L}]$.
By combining Eq. \eqref{eq: gradient_variance} and Eq. \eqref{eq: second_moment}, we can derive the following relationship (see \cite{SM} for detailed calculations):
\begin{align}
    \label{eq: deep_layer_variance}
    \mathrm{Var}[\partial_k \mathcal{L}]\propto \left(\frac{9}{2^5}\right)^{n}\frac{sN_\mathrm{eff}}{nl}.
\end{align}
A key distinction to note in this result is that, contrary to the single-layer case, increasing the number of effective parameters $N_\mathrm{eff}$ for deep circuits increases $\mathrm{Var}[\partial_k \mathcal{L}]$.
Furthermore, we find that $\mathrm{Var}[\partial_k \mathcal{L}]$ is proportional to $\frac{N_\mathrm{eff}}{l}$, which is mentioned in previous study \cite{yao2025directgradientcomputationbarren} for the case where $s=1$.
Upon introducing the $s$-qubit gate, we further find that $\mathrm{Var}[\partial_k \mathcal{L}]$ is proportional to $\frac{s N_\mathrm{eff}}{l}$.
In practice, although employing $s$-qubit gates reduces the number of parameters per layer (from $n$ to $\frac{n}{s}$), the $s  N_\mathrm{eff}$ term in the final result compensates for this reduction, leaving the result of $\mathrm{Var}[\partial_k \mathcal{L}]$ virtually unchanged.
This implies that the scope of the rotation gates within the parameterized layers, whether they are 1-qubit or $s$-qubit gates, does not significantly impact the overall magnitude of the gradient of the cost function.

To empirically validate our theoretical conclusions of Eq.~\eqref{eq: deep_layer_variance}, we performed numerical simulations on quantum circuits with $n=2$ to $n=12$ qubits.
We now proceed to verify our theoretical conclusions with numerical simulations.
We systematically investigated the gradient variance $\mathrm{Var}[\partial_k \mathcal{L}]$ as a function of circuit depth $l$ (from $l=5$ to $150$) and qubit count $n$, considering various observable types and $s$-qubit gate sizes.

\begin{figure}[t]
    \includegraphics[width=\linewidth]{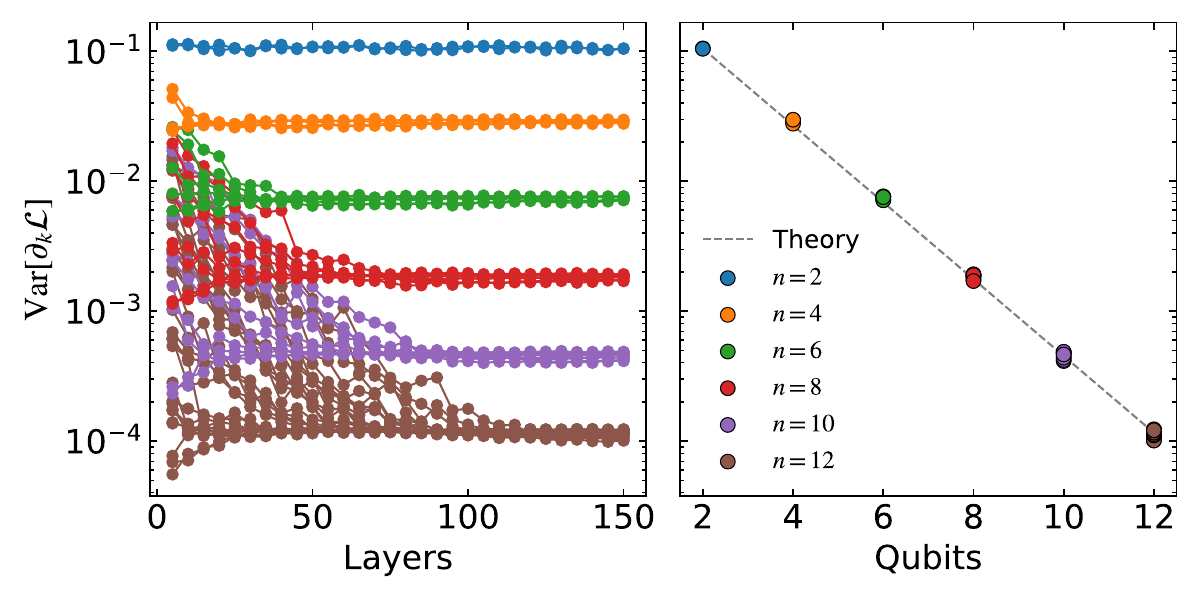}
    \caption{\raggedright Numerical simulation of gradient variance behavior. (Left) Gradient variance $\mathrm{Var}[\partial_k \mathcal{L}]$ as a function of circuit depth $l$ (from 5 to 150) for different numbers of qubits $n$ (from $n=2, 4, \dots, 12$).
    The variance saturates as $l$ increases for all system sizes.
    (Right) Log-scale plot of the gradient variance in the deep layer (at $l=150$) as a function of the number of qubits $n$.
    The clear linear trend (dashed line) indicates an exponential decay of variance with $n$.}
    \label{fig: deep_layer}
\end{figure}

As shown in Figure.~\ref{fig: deep_layer}, we first observe that the gradient variance $\mathrm{Var}[\partial_k \mathcal{L}]$ rapidly converges and saturates to a stable value as the circuit depth $l$ increases.
This saturation behavior is consistent with our theoretical prediction.
As the layer becomes deep enough, the $\frac{sN_\mathrm{eff}}{nl}$ term in Eq.~\eqref{eq: deep_layer_variance} approaches $1$, causing the saturated variance to be primarily dependent on the system size $n$.
To precisely quantify the variance's dependence on $n$, we extracted the saturated variance values at a sufficiently deep level, $l=150$.
In Figure.~\ref{fig: deep_layer}, we plot the logarithm of this saturated variance, $\log(\mathrm{Var}[\partial_k \mathcal{L}])\propto O(n)$, against the number of qubits $n$, which is mentioned in previous studies \cite{McClean2018, yao2025directgradientcomputationbarren}.
The data points exhibit a clear linear trend, providing strong numerical evidence that the gradient variance decays exponentially with the system size $n$.
The resulting linear fit offers a quantitative description of this exponential relationship.

\begin{figure}[t]
    \centering
    \includegraphics[width=0.9\linewidth]{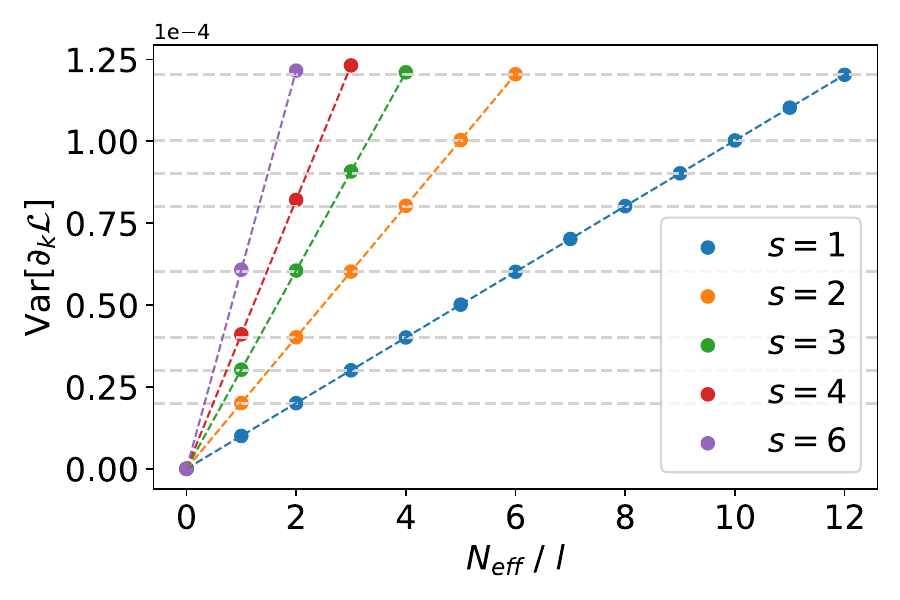}
    \caption{\raggedright Gradient variance $\mathrm{Var}[\partial_k \mathcal{L}]$ as a function of the factor $\frac{s N_\mathrm{eff}} {l}$.
    The plot includes all experimental data for a fixed qubit count $n=12$ and observable $Z^{\otimes 12}$ across various $s$-qubit gate sizes ($s=1, 2, 3, 4, 6$) and different $\frac{N_\mathrm{eff}}{l}$ ratios via pruning. The points on the same gray line mean the same $ sN_\mathrm{eff}$.}
    \label{fig: deep_s}
\end{figure}

We then fixed the number of qubits to $n=12$, observable to $Z^{\otimes 12}$ and employed $s$-qubit gates of varying sizes, specifically $s=1, 2, 3, 4, \text{and } 6$, we can get the result of $\mathrm{Var}[\partial_k \mathcal{L}]\propto\frac{s  N_\mathrm{eff}}{l}$.
It is crucial to note that the choice of $s$ physically constrains the maximum number of parameters a layer can accommodate.
For each $s$ configuration, we therefore precisely controlled $N_\mathrm{eff}$ by randomly pruning a fraction of parameterized gates across the entire circuit, allowing us to get a range of $\frac{N_\mathrm{eff}}{l}$ ratios.
As depicted in Figure.~\ref{fig: deep_s}, we plot all experimental data that encompass all combinations of $s$ and $N_\mathrm{eff}$ on a single graph with $\mathrm{Var}[\partial_k \mathcal{L}]$ on the y-axis and our proposed factor $\frac{N_\mathrm{eff}}{l}$ on the x-axis.
The data in Figure.~\ref{fig: deep_s} reveals this unified scaling in three distinct steps:
\begin{enumerate}
    \item For any fixed $s$, $\mathrm{Var}[\partial_k \mathcal{L}]$ exhibits a clear linear dependence on the $\frac{N_\mathrm{eff}}{l}$ ratio.
    \item Crucially, when comparing across different $s$ configurations, we observe that similar variance magnitudes consistently correspond to nearly identical $\frac{sN_\mathrm{eff}}{l}$ values.
    \item As expected, when $N_\mathrm{eff}=0$, signifying no parameterized gates in the circuit, the variance is precisely zero, anchoring the linear trend at the origin.
\end{enumerate}
This provides compelling evidence that the gradient variance is not governed by $s$ or $\frac{N_\mathrm{eff}}{l}$ in isolation, but is strictly controlled by the unified factor $\frac{sN_\mathrm{eff}}{l}$.
The findings in \cite{yao2025directgradientcomputationbarren} can be seen as the $s=1$ special case of this relationship; our result thus provides a clear generalization.

\textit{Conclusion.---}
In this Letter, we extend a method for gradient computation that avoids the Haar random assumption on parameters \cite{yao2025directgradientcomputationbarren}, expanding the set of Pauli operators in the rotation gates from $\{X, Y, Z\}$ to $\{I, X, Y, Z\}^{\otimes s}$.
This approach provides a more direct perspective on the gradient distribution of the cost function, offering precise insights into the underlying characteristics of gradient behavior.
Crucially, it enables a unified analysis of gradients in both shallow and deep PQCs, overcoming the limitation of traditional Haar random-based methods, which are ill-suited for analyzing shallow circuits.

Based on this framework, we have made two key advances. First, for single-layer PQCs, we derive exact analytical expressions for the gradient expectation and variance for different local generators.
We explain the impact of the number of ``effective parameters" on the gradient variance.
This finding lends theoretical support to the superior performance of local observables in shallow circuits \cite{Cerezo2021}.
Second, for deep-layer PQCs, our work not only verifies the barren plateau phenomenon, where gradient variance vanishes exponentially with the qubit count \cite{McClean2018}, but more significantly, it reveals how the gradient variance is jointly modulated by the number of effective parameters and the local generator.
Our analysis characterizes this dependency, quantifying how the gradient vanishing is governed by four factors: the number of qubits ($n$), layers ($l$), effective parameter($N_\mathrm{eff}$), and the size of the multi-qubit gate ($s$), thus offering new insights into the underlying mechanisms of vanishing gradients in deep circuits.

Our proposed direct analysis framework avoids potential complications arising from complex mathematical concepts (such as Haar random and $t$-design) or ambiguous definitions of unitary operations.
The method is currently limited to PQCs architectures where the type of rotation gates is uniformly sampled from the gate set.
Despite this restriction, this setting is consistent with the random circuit models used in most contemporary numerical simulations.
In the future, we intend to generalize this approach to a broader class of unitary operators and explore its applicability to more complex circuit architectures, such as hardware-efficient ansatze.

\nocite{*}

\textit{acknowledgments.---}
This work was supported by JSPS KAKENHI Grant Number JP23K24915.

% \bibliography{main}% Produces the bibliography via BibTeX.

\begin{thebibliography}{32}%
\makeatletter
\providecommand \@ifxundefined [1]{%
 \@ifx{#1\undefined}
}%
\providecommand \@ifnum [1]{%
 \ifnum #1\expandafter \@firstoftwo
 \else \expandafter \@secondoftwo
 \fi
}%
\providecommand \@ifx [1]{%
 \ifx #1\expandafter \@firstoftwo
 \else \expandafter \@secondoftwo
 \fi
}%
\providecommand \natexlab [1]{#1}%
\providecommand \enquote  [1]{``#1''}%
\providecommand \bibnamefont  [1]{#1}%
\providecommand \bibfnamefont [1]{#1}%
\providecommand \citenamefont [1]{#1}%
\providecommand \href@noop [0]{\@secondoftwo}%
\providecommand \href [0]{\begingroup \@sanitize@url \@href}%
\providecommand \@href[1]{\@@startlink{#1}\@@href}%
\providecommand \@@href[1]{\endgroup#1\@@endlink}%
\providecommand \@sanitize@url [0]{\catcode `\\12\catcode `\$12\catcode `\&12\catcode `\#12\catcode `\^12\catcode `\_12\catcode `\%12\relax}%
\providecommand \@@startlink[1]{}%
\providecommand \@@endlink[0]{}%
\providecommand \url  [0]{\begingroup\@sanitize@url \@url }%
\providecommand \@url [1]{\endgroup\@href {#1}{\urlprefix }}%
\providecommand \urlprefix  [0]{URL }%
\providecommand \Eprint [0]{\href }%
\providecommand \doibase [0]{https://doi.org/}%
\providecommand \selectlanguage [0]{\@gobble}%
\providecommand \bibinfo  [0]{\@secondoftwo}%
\providecommand \bibfield  [0]{\@secondoftwo}%
\providecommand \translation [1]{[#1]}%
\providecommand \BibitemOpen [0]{}%
\providecommand \bibitemStop [0]{}%
\providecommand \bibitemNoStop [0]{.\EOS\space}%
\providecommand \EOS [0]{\spacefactor3000\relax}%
\providecommand \BibitemShut  [1]{\csname bibitem#1\endcsname}%
\let\auto@bib@innerbib\@empty
%</preamble>
\bibitem [{\citenamefont {Yao}\ and\ \citenamefont {Hasegawa}(2025{\natexlab{a}})}]{yao2025directgradientcomputationbarren}%
  \BibitemOpen
  \bibfield  {author} {\bibinfo {author} {\bibfnamefont {Y.}~\bibnamefont {Yao}}\ and\ \bibinfo {author} {\bibfnamefont {Y.}~\bibnamefont {Hasegawa}},\ }\bibfield  {title} {\bibinfo {title} {Linking barren plateaus to effective parameters in deep rotation-gate-based parameterized quantum circuits},\ }\href {https://doi.org/10.1103/bz8d-2yv2} {\bibfield  {journal} {\bibinfo  {journal} {Phys. Rev. A}\ }\textbf {\bibinfo {volume} {112}},\ \bibinfo {pages} {062443} (\bibinfo {year} {2025}{\natexlab{a}})}\BibitemShut {NoStop}%
\bibitem [{\citenamefont {McClean}\ \emph {et~al.}(2018)\citenamefont {McClean}, \citenamefont {Boixo}, \citenamefont {Smelyanskiy}, \citenamefont {Babbush},\ and\ \citenamefont {Neven}}]{McClean2018}%
  \BibitemOpen
  \bibfield  {author} {\bibinfo {author} {\bibfnamefont {J.~R.}\ \bibnamefont {McClean}}, \bibinfo {author} {\bibfnamefont {S.}~\bibnamefont {Boixo}}, \bibinfo {author} {\bibfnamefont {V.~N.}\ \bibnamefont {Smelyanskiy}}, \bibinfo {author} {\bibfnamefont {R.}~\bibnamefont {Babbush}},\ and\ \bibinfo {author} {\bibfnamefont {H.}~\bibnamefont {Neven}},\ }\bibfield  {title} {\bibinfo {title} {Barren plateaus in quantum neural network training landscapes},\ }\href {https://doi.org/10.1038/s41467-018-07090-4} {\bibfield  {journal} {\bibinfo  {journal} {Nature Communications}\ }\textbf {\bibinfo {volume} {9}},\ \bibinfo {pages} {4812} (\bibinfo {year} {2018})}\BibitemShut {NoStop}%
\bibitem [{\citenamefont {Cerezo}\ \emph {et~al.}(2021)\citenamefont {Cerezo}, \citenamefont {Sone}, \citenamefont {Volkoff}, \citenamefont {Cincio},\ and\ \citenamefont {Coles}}]{Cerezo2021}%
  \BibitemOpen
  \bibfield  {author} {\bibinfo {author} {\bibfnamefont {M.}~\bibnamefont {Cerezo}}, \bibinfo {author} {\bibfnamefont {A.}~\bibnamefont {Sone}}, \bibinfo {author} {\bibfnamefont {T.}~\bibnamefont {Volkoff}}, \bibinfo {author} {\bibfnamefont {L.}~\bibnamefont {Cincio}},\ and\ \bibinfo {author} {\bibfnamefont {P.~J.}\ \bibnamefont {Coles}},\ }\bibfield  {title} {\bibinfo {title} {Cost function dependent barren plateaus in shallow parametrized quantum circuits},\ }\href {https://doi.org/10.1038/s41467-021-21728-w} {\bibfield  {journal} {\bibinfo  {journal} {Nature Communications}\ }\textbf {\bibinfo {volume} {12}},\ \bibinfo {pages} {1791} (\bibinfo {year} {2021})}\BibitemShut {NoStop}%
\bibitem [{\citenamefont {Anshu}\ and\ \citenamefont {Metger}(2023)}]{Anshu2023intro}%
  \BibitemOpen
  \bibfield  {author} {\bibinfo {author} {\bibfnamefont {A.}~\bibnamefont {Anshu}}\ and\ \bibinfo {author} {\bibfnamefont {T.}~\bibnamefont {Metger}},\ }\bibfield  {title} {\bibinfo {title} {Concentration bounds for quantum states and limitations on the {QAOA} from polynomial approximations},\ }\href {https://doi.org/10.22331/q-2023-05-11-999} {\bibfield  {journal} {\bibinfo  {journal} {{Quantum}}\ }\textbf {\bibinfo {volume} {7}},\ \bibinfo {pages} {999} (\bibinfo {year} {2023})}\BibitemShut {NoStop}%
\bibitem [{\citenamefont {Leone}\ \emph {et~al.}(2024{\natexlab{a}})\citenamefont {Leone}, \citenamefont {Oliviero}, \citenamefont {Cincio},\ and\ \citenamefont {Cerezo}}]{Leone2024}%
  \BibitemOpen
  \bibfield  {author} {\bibinfo {author} {\bibfnamefont {L.}~\bibnamefont {Leone}}, \bibinfo {author} {\bibfnamefont {S.~F.}\ \bibnamefont {Oliviero}}, \bibinfo {author} {\bibfnamefont {L.}~\bibnamefont {Cincio}},\ and\ \bibinfo {author} {\bibfnamefont {M.}~\bibnamefont {Cerezo}},\ }\bibfield  {title} {\bibinfo {title} {On the practical usefulness of the {H}ardware {E}fficient {A}nsatz},\ }\href {https://doi.org/10.22331/q-2024-07-03-1395} {\bibfield  {journal} {\bibinfo  {journal} {{Quantum}}\ }\textbf {\bibinfo {volume} {8}},\ \bibinfo {pages} {1395} (\bibinfo {year} {2024}{\natexlab{a}})}\BibitemShut {NoStop}%
\bibitem [{\citenamefont {Uvarov}\ and\ \citenamefont {Biamonte}(2021)}]{Uvarov2021gradient}%
  \BibitemOpen
  \bibfield  {author} {\bibinfo {author} {\bibfnamefont {A.~V.}\ \bibnamefont {Uvarov}}\ and\ \bibinfo {author} {\bibfnamefont {J.~D.}\ \bibnamefont {Biamonte}},\ }\bibfield  {title} {\bibinfo {title} {On barren plateaus and cost function locality in variational quantum algorithms},\ }\href {https://doi.org/10.1088/1751-8121/abfac7} {\bibfield  {journal} {\bibinfo  {journal} {Journal of Physics A: Mathematical and Theoretical}\ }\textbf {\bibinfo {volume} {54}},\ \bibinfo {pages} {245301} (\bibinfo {year} {2021})}\BibitemShut {NoStop}%
\bibitem [{\citenamefont {Letcher}\ \emph {et~al.}(2024)\citenamefont {Letcher}, \citenamefont {Woerner},\ and\ \citenamefont {Zoufal}}]{Letcher2024gradient}%
  \BibitemOpen
  \bibfield  {author} {\bibinfo {author} {\bibfnamefont {A.}~\bibnamefont {Letcher}}, \bibinfo {author} {\bibfnamefont {S.}~\bibnamefont {Woerner}},\ and\ \bibinfo {author} {\bibfnamefont {C.}~\bibnamefont {Zoufal}},\ }\bibfield  {title} {\bibinfo {title} {Tight and {E}fficient {G}radient {B}ounds for {P}arameterized {Q}uantum {C}ircuits},\ }\href {https://doi.org/10.22331/q-2024-09-25-1484} {\bibfield  {journal} {\bibinfo  {journal} {{Quantum}}\ }\textbf {\bibinfo {volume} {8}},\ \bibinfo {pages} {1484} (\bibinfo {year} {2024})}\BibitemShut {NoStop}%
\bibitem [{\citenamefont {Napp}(2022)}]{napp2022gradient}%
  \BibitemOpen
  \bibfield  {author} {\bibinfo {author} {\bibfnamefont {J.}~\bibnamefont {Napp}},\ }\href {https://arxiv.org/abs/2203.06174} {\bibinfo {title} {Quantifying the barren plateau phenomenon for a model of unstructured variational ans\"{a}tze}} (\bibinfo {year} {2022}),\ \Eprint {https://arxiv.org/abs/2203.06174} {arXiv:2203.06174 [quant-ph]} \BibitemShut {NoStop}%
\bibitem [{\citenamefont {Ragone}\ \emph {et~al.}(2024)\citenamefont {Ragone}, \citenamefont {Bakalov}, \citenamefont {Sauvage}, \citenamefont {Kemper}, \citenamefont {Marrero}, \citenamefont {Larocca},\ and\ \citenamefont {Cerezo}}]{Ragone2024lie}%
  \BibitemOpen
  \bibfield  {author} {\bibinfo {author} {\bibfnamefont {M.}~\bibnamefont {Ragone}}, \bibinfo {author} {\bibfnamefont {B.~N.}\ \bibnamefont {Bakalov}}, \bibinfo {author} {\bibfnamefont {F.}~\bibnamefont {Sauvage}}, \bibinfo {author} {\bibfnamefont {A.~F.}\ \bibnamefont {Kemper}}, \bibinfo {author} {\bibfnamefont {C.~O.}\ \bibnamefont {Marrero}}, \bibinfo {author} {\bibfnamefont {M.}~\bibnamefont {Larocca}},\ and\ \bibinfo {author} {\bibfnamefont {M.}~\bibnamefont {Cerezo}},\ }\bibfield  {title} {\bibinfo {title} {A lie algebraic theory of barren plateaus for deep parameterized quantum circuits},\ }\href {https://doi.org/10.1038/s41467-024-49909-3} {\bibfield  {journal} {\bibinfo  {journal} {Nature Communications}\ }\textbf {\bibinfo {volume} {15}},\ \bibinfo {pages} {7172} (\bibinfo {year} {2024})}\BibitemShut {NoStop}%
\bibitem [{\citenamefont {Holmes}\ \emph {et~al.}(2021)\citenamefont {Holmes}, \citenamefont {Arrasmith}, \citenamefont {Yan}, \citenamefont {Coles}, \citenamefont {Albrecht},\ and\ \citenamefont {Sornborger}}]{holmes_bp}%
  \BibitemOpen
  \bibfield  {author} {\bibinfo {author} {\bibfnamefont {Z.}~\bibnamefont {Holmes}}, \bibinfo {author} {\bibfnamefont {A.}~\bibnamefont {Arrasmith}}, \bibinfo {author} {\bibfnamefont {B.}~\bibnamefont {Yan}}, \bibinfo {author} {\bibfnamefont {P.~J.}\ \bibnamefont {Coles}}, \bibinfo {author} {\bibfnamefont {A.}~\bibnamefont {Albrecht}},\ and\ \bibinfo {author} {\bibfnamefont {A.~T.}\ \bibnamefont {Sornborger}},\ }\bibfield  {title} {\bibinfo {title} {Barren plateaus preclude learning scramblers},\ }\href {https://doi.org/10.1103/PhysRevLett.126.190501} {\bibfield  {journal} {\bibinfo  {journal} {Phys. Rev. Lett.}\ }\textbf {\bibinfo {volume} {126}},\ \bibinfo {pages} {190501} (\bibinfo {year} {2021})}\BibitemShut {NoStop}%
\bibitem [{\citenamefont {Larocca}\ \emph {et~al.}(2022{\natexlab{a}})\citenamefont {Larocca}, \citenamefont {Czarnik}, \citenamefont {Sharma}, \citenamefont {Muraleedharan}, \citenamefont {Coles},\ and\ \citenamefont {Cerezo}}]{Larocca_bp}%
  \BibitemOpen
  \bibfield  {author} {\bibinfo {author} {\bibfnamefont {M.}~\bibnamefont {Larocca}}, \bibinfo {author} {\bibfnamefont {P.}~\bibnamefont {Czarnik}}, \bibinfo {author} {\bibfnamefont {K.}~\bibnamefont {Sharma}}, \bibinfo {author} {\bibfnamefont {G.}~\bibnamefont {Muraleedharan}}, \bibinfo {author} {\bibfnamefont {P.~J.}\ \bibnamefont {Coles}},\ and\ \bibinfo {author} {\bibfnamefont {M.}~\bibnamefont {Cerezo}},\ }\bibfield  {title} {\bibinfo {title} {Diagnosing {B}arren {P}lateaus with {T}ools from {Q}uantum {O}ptimal {C}ontrol},\ }\href {https://doi.org/10.22331/q-2022-09-29-824} {\bibfield  {journal} {\bibinfo  {journal} {{Quantum}}\ }\textbf {\bibinfo {volume} {6}},\ \bibinfo {pages} {824} (\bibinfo {year} {2022}{\natexlab{a}})}\BibitemShut {NoStop}%
\bibitem [{\citenamefont {Sack}\ \emph {et~al.}(2022)\citenamefont {Sack}, \citenamefont {Medina}, \citenamefont {Michailidis}, \citenamefont {Kueng},\ and\ \citenamefont {Serbyn}}]{sack_intro}%
  \BibitemOpen
  \bibfield  {author} {\bibinfo {author} {\bibfnamefont {S.~H.}\ \bibnamefont {Sack}}, \bibinfo {author} {\bibfnamefont {R.~A.}\ \bibnamefont {Medina}}, \bibinfo {author} {\bibfnamefont {A.~A.}\ \bibnamefont {Michailidis}}, \bibinfo {author} {\bibfnamefont {R.}~\bibnamefont {Kueng}},\ and\ \bibinfo {author} {\bibfnamefont {M.}~\bibnamefont {Serbyn}},\ }\bibfield  {title} {\bibinfo {title} {Avoiding barren plateaus using classical shadows},\ }\href {https://doi.org/10.1103/PRXQuantum.3.020365} {\bibfield  {journal} {\bibinfo  {journal} {PRX Quantum}\ }\textbf {\bibinfo {volume} {3}},\ \bibinfo {pages} {020365} (\bibinfo {year} {2022})}\BibitemShut {NoStop}%
\bibitem [{\citenamefont {Du}\ \emph {et~al.}(2022)\citenamefont {Du}, \citenamefont {Tu}, \citenamefont {Yuan},\ and\ \citenamefont {Tao}}]{du2022intro}%
  \BibitemOpen
  \bibfield  {author} {\bibinfo {author} {\bibfnamefont {Y.}~\bibnamefont {Du}}, \bibinfo {author} {\bibfnamefont {Z.}~\bibnamefont {Tu}}, \bibinfo {author} {\bibfnamefont {X.}~\bibnamefont {Yuan}},\ and\ \bibinfo {author} {\bibfnamefont {D.}~\bibnamefont {Tao}},\ }\bibfield  {title} {\bibinfo {title} {Efficient measure for the expressivity of variational quantum algorithms},\ }\href {https://journals.aps.org/prl/abstract/10.1103/PhysRevLett.128.080506} {\bibfield  {journal} {\bibinfo  {journal} {Phys. Rev. Lett.}\ }\textbf {\bibinfo {volume} {128}},\ \bibinfo {pages} {080506} (\bibinfo {year} {2022})}\BibitemShut {NoStop}%
\bibitem [{\citenamefont {Shen}\ \emph {et~al.}(2020)\citenamefont {Shen}, \citenamefont {Zhang}, \citenamefont {You},\ and\ \citenamefont {Zhai}}]{shen2020intro}%
  \BibitemOpen
  \bibfield  {author} {\bibinfo {author} {\bibfnamefont {H.}~\bibnamefont {Shen}}, \bibinfo {author} {\bibfnamefont {P.}~\bibnamefont {Zhang}}, \bibinfo {author} {\bibfnamefont {Y.-Z.}\ \bibnamefont {You}},\ and\ \bibinfo {author} {\bibfnamefont {H.}~\bibnamefont {Zhai}},\ }\bibfield  {title} {\bibinfo {title} {Information scrambling in quantum neural networks},\ }\href {https://journals.aps.org/prl/abstract/10.1103/PhysRevLett.124.200504} {\bibfield  {journal} {\bibinfo  {journal} {Phys. Rev. Lett.}\ }\textbf {\bibinfo {volume} {124}},\ \bibinfo {pages} {200504} (\bibinfo {year} {2020})}\BibitemShut {NoStop}%
\bibitem [{\citenamefont {Patti}\ \emph {et~al.}(2021)\citenamefont {Patti}, \citenamefont {Najafi}, \citenamefont {Gao},\ and\ \citenamefont {Yelin}}]{patti2021intro}%
  \BibitemOpen
  \bibfield  {author} {\bibinfo {author} {\bibfnamefont {T.~L.}\ \bibnamefont {Patti}}, \bibinfo {author} {\bibfnamefont {K.}~\bibnamefont {Najafi}}, \bibinfo {author} {\bibfnamefont {X.}~\bibnamefont {Gao}},\ and\ \bibinfo {author} {\bibfnamefont {S.~F.}\ \bibnamefont {Yelin}},\ }\bibfield  {title} {\bibinfo {title} {Entanglement devised barren plateau mitigation},\ }\href {https://journals.aps.org/prresearch/abstract/10.1103/PhysRevResearch.3.033090} {\bibfield  {journal} {\bibinfo  {journal} {Physical Review Research}\ }\textbf {\bibinfo {volume} {3}},\ \bibinfo {pages} {033090} (\bibinfo {year} {2021})}\BibitemShut {NoStop}%
\bibitem [{\citenamefont {Wurtz}\ and\ \citenamefont {Love}(2021)}]{wurtz2021intro}%
  \BibitemOpen
  \bibfield  {author} {\bibinfo {author} {\bibfnamefont {J.}~\bibnamefont {Wurtz}}\ and\ \bibinfo {author} {\bibfnamefont {P.}~\bibnamefont {Love}},\ }\bibfield  {title} {\bibinfo {title} {Maxcut quantum approximate optimization algorithm performance guarantees for p> 1},\ }\href@noop {} {\bibfield  {journal} {\bibinfo  {journal} {Phys. Rev. A}\ }\textbf {\bibinfo {volume} {103}},\ \bibinfo {pages} {042612} (\bibinfo {year} {2021})}\BibitemShut {NoStop}%
\bibitem [{\citenamefont {Larocca}\ \emph {et~al.}(2022{\natexlab{b}})\citenamefont {Larocca}, \citenamefont {Czarnik}, \citenamefont {Sharma}, \citenamefont {Muraleedharan}, \citenamefont {Coles},\ and\ \citenamefont {Cerezo}}]{larocca2022intro}%
  \BibitemOpen
  \bibfield  {author} {\bibinfo {author} {\bibfnamefont {M.}~\bibnamefont {Larocca}}, \bibinfo {author} {\bibfnamefont {P.}~\bibnamefont {Czarnik}}, \bibinfo {author} {\bibfnamefont {K.}~\bibnamefont {Sharma}}, \bibinfo {author} {\bibfnamefont {G.}~\bibnamefont {Muraleedharan}}, \bibinfo {author} {\bibfnamefont {P.~J.}\ \bibnamefont {Coles}},\ and\ \bibinfo {author} {\bibfnamefont {M.}~\bibnamefont {Cerezo}},\ }\bibfield  {title} {\bibinfo {title} {Diagnosing barren plateaus with tools from quantum optimal control},\ }\href@noop {} {\bibfield  {journal} {\bibinfo  {journal} {Quantum}\ }\textbf {\bibinfo {volume} {6}},\ \bibinfo {pages} {824} (\bibinfo {year} {2022}{\natexlab{b}})}\BibitemShut {NoStop}%
\bibitem [{\citenamefont {Russell}\ \emph {et~al.}(2017)\citenamefont {Russell}, \citenamefont {Rabitz},\ and\ \citenamefont {Wu}}]{Russell2017intro}%
  \BibitemOpen
  \bibfield  {author} {\bibinfo {author} {\bibfnamefont {B.}~\bibnamefont {Russell}}, \bibinfo {author} {\bibfnamefont {H.}~\bibnamefont {Rabitz}},\ and\ \bibinfo {author} {\bibfnamefont {R.-B.}\ \bibnamefont {Wu}},\ }\bibfield  {title} {\bibinfo {title} {Control landscapes are almost always trap free: a geometric assessment},\ }\href {https://doi.org/10.1088/1751-8121/aa6b77} {\bibfield  {journal} {\bibinfo  {journal} {Journal of Physics A: Mathematical and Theoretical}\ }\textbf {\bibinfo {volume} {50}},\ \bibinfo {pages} {205302} (\bibinfo {year} {2017})}\BibitemShut {NoStop}%
\bibitem [{\citenamefont {Wiersema}\ \emph {et~al.}(2020)\citenamefont {Wiersema}, \citenamefont {Zhou}, \citenamefont {de~Sereville}, \citenamefont {Carrasquilla}, \citenamefont {Kim},\ and\ \citenamefont {Yuen}}]{wiersema2020intro}%
  \BibitemOpen
  \bibfield  {author} {\bibinfo {author} {\bibfnamefont {R.}~\bibnamefont {Wiersema}}, \bibinfo {author} {\bibfnamefont {C.}~\bibnamefont {Zhou}}, \bibinfo {author} {\bibfnamefont {Y.}~\bibnamefont {de~Sereville}}, \bibinfo {author} {\bibfnamefont {J.~F.}\ \bibnamefont {Carrasquilla}}, \bibinfo {author} {\bibfnamefont {Y.~B.}\ \bibnamefont {Kim}},\ and\ \bibinfo {author} {\bibfnamefont {H.}~\bibnamefont {Yuen}},\ }\bibfield  {title} {\bibinfo {title} {Exploring entanglement and optimization within the hamiltonian variational ansatz},\ }\href {https://doi.org/10.1103/PRXQuantum.1.020319} {\bibfield  {journal} {\bibinfo  {journal} {PRX Quantum}\ }\textbf {\bibinfo {volume} {1}},\ \bibinfo {pages} {020319} (\bibinfo {year} {2020})}\BibitemShut {NoStop}%
\bibitem [{\citenamefont {Sciorilli}\ \emph {et~al.}(2025)\citenamefont {Sciorilli}, \citenamefont {Borges}, \citenamefont {Patti}, \citenamefont {Garc\'{i}a-Mart\'{i}n}, \citenamefont {Camilo}, \citenamefont {Anandkumar},\ and\ \citenamefont {Aolita}}]{Sciorilli2025intro}%
  \BibitemOpen
  \bibfield  {author} {\bibinfo {author} {\bibfnamefont {M.}~\bibnamefont {Sciorilli}}, \bibinfo {author} {\bibfnamefont {L.}~\bibnamefont {Borges}}, \bibinfo {author} {\bibfnamefont {T.~L.}\ \bibnamefont {Patti}}, \bibinfo {author} {\bibfnamefont {D.}~\bibnamefont {Garc\'{i}a-Mart\'{i}n}}, \bibinfo {author} {\bibfnamefont {G.}~\bibnamefont {Camilo}}, \bibinfo {author} {\bibfnamefont {A.}~\bibnamefont {Anandkumar}},\ and\ \bibinfo {author} {\bibfnamefont {L.}~\bibnamefont {Aolita}},\ }\bibfield  {title} {\bibinfo {title} {Towards large-scale quantum optimization solvers with few qubits},\ }\href {https://doi.org/10.1038/s41467-024-55346-z} {\bibfield  {journal} {\bibinfo  {journal} {Nature Communications}\ }\textbf {\bibinfo {volume} {16}},\ \bibinfo {pages} {476} (\bibinfo {year} {2025})}\BibitemShut {NoStop}%
\bibitem [{\citenamefont {Wild}\ and\ \citenamefont {Alhambra}(2023)}]{Wild2023intro}%
  \BibitemOpen
  \bibfield  {author} {\bibinfo {author} {\bibfnamefont {D.~S.}\ \bibnamefont {Wild}}\ and\ \bibinfo {author} {\bibfnamefont {A.~M.}\ \bibnamefont {Alhambra}},\ }\bibfield  {title} {\bibinfo {title} {Classical simulation of short-time quantum dynamics},\ }\href {https://doi.org/10.1103/PRXQuantum.4.020340} {\bibfield  {journal} {\bibinfo  {journal} {PRX Quantum}\ }\textbf {\bibinfo {volume} {4}},\ \bibinfo {pages} {020340} (\bibinfo {year} {2023})}\BibitemShut {NoStop}%
\bibitem [{\citenamefont {De~Palma}\ \emph {et~al.}(2023)\citenamefont {De~Palma}, \citenamefont {Marvian}, \citenamefont {Rouz\'e},\ and\ \citenamefont {Fran\ifmmode~\mbox{\c{c}}\else \c{c}\fi{}a}}]{Palma2023intro}%
  \BibitemOpen
  \bibfield  {author} {\bibinfo {author} {\bibfnamefont {G.}~\bibnamefont {De~Palma}}, \bibinfo {author} {\bibfnamefont {M.}~\bibnamefont {Marvian}}, \bibinfo {author} {\bibfnamefont {C.}~\bibnamefont {Rouz\'e}},\ and\ \bibinfo {author} {\bibfnamefont {D.~S.}\ \bibnamefont {Fran\ifmmode~\mbox{\c{c}}\else \c{c}\fi{}a}},\ }\bibfield  {title} {\bibinfo {title} {Limitations of variational quantum algorithms: A quantum optimal transport approach},\ }\href {https://doi.org/10.1103/PRXQuantum.4.010309} {\bibfield  {journal} {\bibinfo  {journal} {PRX Quantum}\ }\textbf {\bibinfo {volume} {4}},\ \bibinfo {pages} {010309} (\bibinfo {year} {2023})}\BibitemShut {NoStop}%
\bibitem [{\citenamefont {Caro}\ \emph {et~al.}(2023)\citenamefont {Caro}, \citenamefont {Huang}, \citenamefont {Ezzell}, \citenamefont {Gibbs}, \citenamefont {Sornborger}, \citenamefont {Cincio}, \citenamefont {Coles},\ and\ \citenamefont {Holmes}}]{Caro2023intro}%
  \BibitemOpen
  \bibfield  {author} {\bibinfo {author} {\bibfnamefont {M.~C.}\ \bibnamefont {Caro}}, \bibinfo {author} {\bibfnamefont {H.~Y.}\ \bibnamefont {Huang}}, \bibinfo {author} {\bibfnamefont {N.}~\bibnamefont {Ezzell}}, \bibinfo {author} {\bibfnamefont {J.}~\bibnamefont {Gibbs}}, \bibinfo {author} {\bibfnamefont {A.~T.}\ \bibnamefont {Sornborger}}, \bibinfo {author} {\bibfnamefont {L.}~\bibnamefont {Cincio}}, \bibinfo {author} {\bibfnamefont {P.~J.}\ \bibnamefont {Coles}},\ and\ \bibinfo {author} {\bibfnamefont {Z.}~\bibnamefont {Holmes}},\ }\bibfield  {title} {\bibinfo {title} {Out-of-distribution generalization for learning quantum dynamics},\ }\href {https://doi.org/10.1038/s41467-023-39381-w} {\bibfield  {journal} {\bibinfo  {journal} {Nature Communications}\ }\textbf {\bibinfo {volume} {14}},\ \bibinfo {pages} {3751} (\bibinfo {year} {2023})}\BibitemShut {NoStop}%
\bibitem [{\citenamefont {Jerbi}\ \emph {et~al.}(2023)\citenamefont {Jerbi}, \citenamefont {Fiderer}, \citenamefont {Nautrup}, \citenamefont {K\"ubler}, \citenamefont {Briegel},\ and\ \citenamefont {Dunjko}}]{Jerbi2023intro}%
  \BibitemOpen
  \bibfield  {author} {\bibinfo {author} {\bibfnamefont {S.}~\bibnamefont {Jerbi}}, \bibinfo {author} {\bibfnamefont {L.~J.}\ \bibnamefont {Fiderer}}, \bibinfo {author} {\bibfnamefont {H.~P.}\ \bibnamefont {Nautrup}}, \bibinfo {author} {\bibfnamefont {J.~M.}\ \bibnamefont {K\"ubler}}, \bibinfo {author} {\bibfnamefont {H.~J.}\ \bibnamefont {Briegel}},\ and\ \bibinfo {author} {\bibfnamefont {V.}~\bibnamefont {Dunjko}},\ }\bibfield  {title} {\bibinfo {title} {Quantum machine learning beyond kernel methods},\ }\href {https://doi.org/10.1038/s41467-023-36159-y} {\bibfield  {journal} {\bibinfo  {journal} {Nature Communications}\ }\textbf {\bibinfo {volume} {14}},\ \bibinfo {pages} {517} (\bibinfo {year} {2023})}\BibitemShut {NoStop}%
\bibitem [{\citenamefont {Wiersema}\ \emph {et~al.}(2024)\citenamefont {Wiersema}, \citenamefont {K\"okc\"u}, \citenamefont {Kemper},\ and\ \citenamefont {Bakalov}}]{Wiersema2024intro}%
  \BibitemOpen
  \bibfield  {author} {\bibinfo {author} {\bibfnamefont {R.}~\bibnamefont {Wiersema}}, \bibinfo {author} {\bibfnamefont {E.}~\bibnamefont {K\"okc\"u}}, \bibinfo {author} {\bibfnamefont {A.~F.}\ \bibnamefont {Kemper}},\ and\ \bibinfo {author} {\bibfnamefont {B.~N.}\ \bibnamefont {Bakalov}},\ }\bibfield  {title} {\bibinfo {title} {Classification of dynamical lie algebras of 2-local spin systems on linear, circular and fully connected topologies},\ }\href {https://doi.org/10.1038/s41534-024-00900-2} {\bibfield  {journal} {\bibinfo  {journal} {npj Quantum Information}\ }\textbf {\bibinfo {volume} {10}},\ \bibinfo {pages} {110} (\bibinfo {year} {2024})}\BibitemShut {NoStop}%
\bibitem [{\citenamefont {Grant}\ \emph {et~al.}(2019)\citenamefont {Grant}, \citenamefont {Wossnig}, \citenamefont {Ostaszewski},\ and\ \citenamefont {Benedetti}}]{Grant_bp}%
  \BibitemOpen
  \bibfield  {author} {\bibinfo {author} {\bibfnamefont {E.}~\bibnamefont {Grant}}, \bibinfo {author} {\bibfnamefont {L.}~\bibnamefont {Wossnig}}, \bibinfo {author} {\bibfnamefont {M.}~\bibnamefont {Ostaszewski}},\ and\ \bibinfo {author} {\bibfnamefont {M.}~\bibnamefont {Benedetti}},\ }\bibfield  {title} {\bibinfo {title} {An initialization strategy for addressing barren plateaus in parametrized quantum circuits},\ }\href {https://doi.org/10.22331/q-2019-12-09-214} {\bibfield  {journal} {\bibinfo  {journal} {{Quantum}}\ }\textbf {\bibinfo {volume} {3}},\ \bibinfo {pages} {214} (\bibinfo {year} {2019})}\BibitemShut {NoStop}%
\bibitem [{\citenamefont {Yao}\ and\ \citenamefont {Hasegawa}(2025{\natexlab{b}})}]{yao2025}%
  \BibitemOpen
  \bibfield  {author} {\bibinfo {author} {\bibfnamefont {Y.}~\bibnamefont {Yao}}\ and\ \bibinfo {author} {\bibfnamefont {Y.}~\bibnamefont {Hasegawa}},\ }\bibfield  {title} {\bibinfo {title} {Avoiding barren plateaus with entanglement},\ }\href {https://doi.org/10.1103/PhysRevA.111.022426} {\bibfield  {journal} {\bibinfo  {journal} {Phys. Rev. A}\ }\textbf {\bibinfo {volume} {111}},\ \bibinfo {pages} {022426} (\bibinfo {year} {2025}{\natexlab{b}})}\BibitemShut {NoStop}%
\bibitem [{\citenamefont {Cerezo}\ \emph {et~al.}(2024)\citenamefont {Cerezo}, \citenamefont {Larocca}, \citenamefont {García-Martín}, \citenamefont {Diaz}, \citenamefont {Braccia}, \citenamefont {Fontana}, \citenamefont {Rudolph}, \citenamefont {Bermejo}, \citenamefont {Ijaz}, \citenamefont {Thanasilp}, \citenamefont {Anschuetz},\ and\ \citenamefont {Holmes}}]{cerezo2024structure}%
  \BibitemOpen
  \bibfield  {author} {\bibinfo {author} {\bibfnamefont {M.}~\bibnamefont {Cerezo}}, \bibinfo {author} {\bibfnamefont {M.}~\bibnamefont {Larocca}}, \bibinfo {author} {\bibfnamefont {D.}~\bibnamefont {García-Martín}}, \bibinfo {author} {\bibfnamefont {N.~L.}\ \bibnamefont {Diaz}}, \bibinfo {author} {\bibfnamefont {P.}~\bibnamefont {Braccia}}, \bibinfo {author} {\bibfnamefont {E.}~\bibnamefont {Fontana}}, \bibinfo {author} {\bibfnamefont {M.~S.}\ \bibnamefont {Rudolph}}, \bibinfo {author} {\bibfnamefont {P.}~\bibnamefont {Bermejo}}, \bibinfo {author} {\bibfnamefont {A.}~\bibnamefont {Ijaz}}, \bibinfo {author} {\bibfnamefont {S.}~\bibnamefont {Thanasilp}}, \bibinfo {author} {\bibfnamefont {E.~R.}\ \bibnamefont {Anschuetz}},\ and\ \bibinfo {author} {\bibfnamefont {Z.}~\bibnamefont {Holmes}},\ }\href {https://arxiv.org/abs/2312.09121} {\bibinfo {title} {Does provable absence of barren plateaus imply classical simulability? or, why we need to rethink variational quantum computing}} (\bibinfo {year} {2024}),\
  \Eprint {https://arxiv.org/abs/2312.09121} {arXiv:2312.09121 [quant-ph]} \BibitemShut {NoStop}%
\bibitem [{\citenamefont {Leone}\ \emph {et~al.}(2024{\natexlab{b}})\citenamefont {Leone}, \citenamefont {Oliviero}, \citenamefont {Cincio},\ and\ \citenamefont {Cerezo}}]{Leone2024structure}%
  \BibitemOpen
  \bibfield  {author} {\bibinfo {author} {\bibfnamefont {L.}~\bibnamefont {Leone}}, \bibinfo {author} {\bibfnamefont {S.~F.}\ \bibnamefont {Oliviero}}, \bibinfo {author} {\bibfnamefont {L.}~\bibnamefont {Cincio}},\ and\ \bibinfo {author} {\bibfnamefont {M.}~\bibnamefont {Cerezo}},\ }\bibfield  {title} {\bibinfo {title} {On the practical usefulness of the {H}ardware {E}fficient {A}nsatz},\ }\href {https://doi.org/10.22331/q-2024-07-03-1395} {\bibfield  {journal} {\bibinfo  {journal} {{Quantum}}\ }\textbf {\bibinfo {volume} {8}},\ \bibinfo {pages} {1395} (\bibinfo {year} {2024}{\natexlab{b}})}\BibitemShut {NoStop}%
\bibitem [{\citenamefont {{Supplemental Material}}()}]{SM}%
  \BibitemOpen
  \bibfield  {author} {\bibinfo {author} {\bibnamefont {{Supplemental Material}}},\ }\href@noop {} {\bibinfo {title} {Supplemental material}},\ \bibinfo {howpublished} {https://journals.aps.org/authors/supplemental-material-instructions},\ \bibinfo {note} {see Supplemental Material for the detailed derivation.}\BibitemShut {Stop}%
\bibitem [{\citenamefont {Collins}\ \emph {et~al.}(2022)\citenamefont {Collins}, \citenamefont {Matsumoto},\ and\ \citenamefont {Novak}}]{Collins_2022}%
  \BibitemOpen
  \bibfield  {author} {\bibinfo {author} {\bibfnamefont {B.}~\bibnamefont {Collins}}, \bibinfo {author} {\bibfnamefont {S.}~\bibnamefont {Matsumoto}},\ and\ \bibinfo {author} {\bibfnamefont {J.}~\bibnamefont {Novak}},\ }\bibfield  {title} {\bibinfo {title} {The weingarten calculus},\ }\href {https://doi.org/10.1090/noti2474} {\bibfield  {journal} {\bibinfo  {journal} {Notices of the American Mathematical Society}\ }\textbf {\bibinfo {volume} {69}},\ \bibinfo {pages} {1} (\bibinfo {year} {2022})}\BibitemShut {NoStop}%
\bibitem [{\citenamefont {Weingarten}(1978)}]{weingarten1978}%
  \BibitemOpen
  \bibfield  {author} {\bibinfo {author} {\bibfnamefont {D.}~\bibnamefont {Weingarten}},\ }\bibfield  {title} {\bibinfo {title} {Asymptotic behavior of group integrals in the limit of infinite rank},\ }\href {https://pubs.aip.org/aip/jmp/article/19/5/999/460197/Asymptotic-behavior-of-group-integrals-in-the} {\bibfield  {journal} {\bibinfo  {journal} {Journal of Mathematical Physics}\ }\textbf {\bibinfo {volume} {19}},\ \bibinfo {pages} {999} (\bibinfo {year} {1978})}\BibitemShut {NoStop}%
\end{thebibliography}

\onecolumngrid
\section*{End Matter}

\textit{Appendix A.---}
The circuit parameters can be categorized by their impact on gradient variance.
We define effective parameters as those that yield a non-zero gradient variance.
In contrast, ineffective parameters are characterized by a gradient variance of zero, which leads to the gradient variance vanishing.

In practice, we observe that ineffective parameters are structurally disconnected from the target observable.
Consequently, they do not contribute to the training.

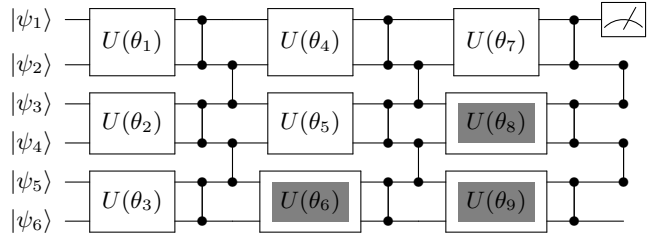
\begin{figure}[h]
    \centering
    \[
\Qcircuit @C=1em @R=.7em {
  % Line 1
  \lstick{\ket{\psi_1}} & \multigate{1}{U(\theta_{1})} & \ctrl{1} & \qw & \multigate{1}{U(\theta_{4})} & \ctrl{1} & \qw & \multigate{1}{U(\theta_{7})} & \ctrl{1} & \meter\\
  % Line 2
  \lstick{\ket{\psi_2}} & \ghost{U(\theta_{1})} & \ctrl{-1} & \ctrl{1} &\ghost{U(\theta_{4})} & \ctrl{-1} & \ctrl{1} & \ghost{U(\theta_{7})} & \ctrl{-1} & \ctrl{1} \\
  % Line 3
  \lstick{\ket{\psi_3}} & \multigate{1}{U(\theta_{2})} & \ctrl{1} & \ctrl{-1} & \multigate{1}{U(\theta_{5})} & \ctrl{1} & \ctrl{-1} & \multigate{1}{{\colorbox{gray}{$U(\theta_{8})$}}} & \ctrl{1} & \ctrl{-1} \\
  % Line 4
  \lstick{\ket{\psi_4}} & \ghost{U(\theta_{2})} & \ctrl{-1} & \ctrl{1} & \ghost{U(\theta_{5})} & \ctrl{-1} & \ctrl{1} & \ghost{{\colorbox{gray}{$U(\theta_{8})$}}} & \ctrl{-1} & \ctrl{1} \\
  \lstick{\ket{\psi_5}} & \multigate{1}{U(\theta_{3})} & \ctrl{1} & \ctrl{-1} & \multigate{1}{{\colorbox{gray}{$U(\theta_{6})$}}} & \ctrl{1} & \ctrl{-1} & \multigate{1}{{\colorbox{gray}{$U(\theta_{9})$}}} & \ctrl{1} & \ctrl{-1} \\
  \lstick{\ket{\psi_6}} & \ghost{U(\theta_{3})} & \ctrl{-1} & \qw & \ghost{{\colorbox{gray}{$U(\theta_{6})$}}} & \ctrl{-1} & \qw & \ghost{{\colorbox{gray}{$U(\theta_{9})$}}} & \ctrl{-1} & \qw 
}
    \]
    \caption{\raggedright Example of the effective parameter. This PQCs consists of a $6$-qubit, $3$-layer circuit, and the size of the multi-qubit gate is $2$. The gray gates, despite containing trainable parameters, do not affect the final measurement output. This is typically because their scope of action does not affect the final observable, and therefore, they have no impact on the loss function's result. This phenomenon demonstrates that not all trainable parameters are effective for a given task.}
    \label{fig: effective_parameters_example}
\end{figure}

\textit{Appendix B.---}
By invoking Eq.~\eqref{eq: rotation_gate_define}, we decompose the second moment of $U(\theta)$ into terms involving the first and second moments of Pauli operators.
Applying the integration results $\cos^4\frac{\theta}{2}=\sin^4\frac{\theta}{2}=\frac{3}{8}$, $\cos^3\frac{\theta}{2}\sin\frac{\theta}{2}=\cos\frac{\theta}{2}\sin^3\frac{\theta}{2}=0$ and $\cos^2\frac{\theta}{2}\sin^2\frac{\theta}{2}=\frac{1}{8}$ (see \cite{SM}), we obtain:
\begin{align}
&\mathrm{E}\left[U(\theta)^\dagger A U(\theta)BU(\theta)^\dagger CU(\theta) \right]\notag\\
=&\mathrm{E}_{\theta,P}\left[\left(\cos\frac{\theta}{2}I+i\sin\frac{\theta}{2}P\right) A\left(\cos\frac{\theta}{2}I-i\sin\frac{\theta}{2}P\right)B\left(\cos\frac{\theta}{2}I+i\sin\frac{\theta}{2}P\right) C\left(\cos\frac{\theta}{2}I-i\sin\frac{\theta}{2}P\right)\right]\notag\\
=&\mathrm{E}_{\theta,P}\left[\cos^4\frac{\theta}{2}ABC+\sin^4\frac{\theta}{2}PAPBPCP+\sin^2\frac{\theta}{2}\cos^2\frac{\theta}{2}\left[PAPBC+ABPCP-\left(PA-AP\right)B\left(PC-CP\right)\right]\right]\notag\\
=&\mathrm{E}_{P}\left[\frac{3}{8}ABC+\frac{3}{8}PAPBPCP+\frac{1}{8}\left(PAPBC+ABPCP+APBPC+PABCP-PABPC-APBCP\right)\right]\notag\\
=&\frac{3}{8}ABC+\frac{3}{8|P|}\sum_PPAPBPCP+\frac{1}{8|P|}\sum_{P}PAPBC+ABPCP+APBPC+PABCP-PABPC-APBCP.
\end{align}

\textit{Appendix C.---}
In the variance expression $\mathrm{Var}[\partial_k \mathcal{L}] = \frac{sN_\mathrm{eff}}{n} F(P) G(P)^{N_{\mathrm{eff}}-1}$, the functions $F(P)$ and $G(P)$ represent the decomposed contributions from the target qubit and the environment qubits, respectively.
Specifically, $F(P)$ characterizes the local variance contribution from the $k$-th qubit where the derivative is taken.
Due to the parameter-shift logic, this qubit undergoes a different operator transformation compared to others, leading to a unique functional form.
For instance, when $P \in \{I, X, Y, Z\}^{\otimes s} \setminus \{I^{\otimes s}\}$, we analytically find $F(P) = \frac{1}{4s} \cdot (\frac{5}{12})^{s-1}$, reflecting the local complexity of the generator $P$.
Conversely, $G(P)$ accounts for the statistical contribution of each remaining $N_{\mathrm{eff}}-1$ effective qubit in the single-layer PQCs.
Due to the structural symmetry of single-layer PQCs and the independent, identical distribution of their parameters, each environment qubit contributes an identical multiplicative factor $G(P)$ to the total variance.
This power-law dependence $G(P)^{N_{\mathrm{eff}}-1}$ directly governs the scaling behavior of the gradient; it provides a rigorous mathematical origin for the barren plateau phenomenon in single PQCs when $G(P) < 1$.
Tab.\ref{tab: single_layer} lists the values of $F(P)$ and $G(P)$ corresponding to different combinations of $P$.
Detailed derivations are provided in the Supplementary Materials \cite{SM}.

\begin{table}[htbp]
    \centering
    \renewcommand{\arraystretch}{1.8}
    \begin{tabular}{lcc}
        \toprule
         & $F(P)$ & $G(P)$ \\
        \midrule
        $P \in \{X, Y, Z\}$ & $\frac{1}{4}$ & $\frac{5}{12}$ \\
        $P \in \{I, X, Y, Z\}^{\otimes 2} / I^{\otimes 2}$ & $\frac{5}{96}$ & $\frac{1}{3}$ \\
        $P \in \{I, X, Y, Z\}^{\otimes 3} / I^{\otimes 3}$ & $\frac{25}{1728}$ & $\frac{17}{126}$ \\
        $P \in \{I, X, Y, Z\}^{\otimes 4} / I^{\otimes 4}$ & $\frac{125}{27648}$ & $\frac{37}{510}$ \\
        \bottomrule
    \end{tabular}
    \caption{The values of $F(P)$ and $G(P)$ under different assembles of $P$.}
    \label{tab: single_layer}
\end{table}

\end{document}